# Electrical Impedance Tomography Based Closed-loop Tumor Treating Fields in Dynamic Lung Tumors

Minmin Wang, *Member*, IEEE, Xu Xie, Yuxi Guo, Liying Zhu, Yue Lan, Haitang Yang, Yun Pan, Guangdi Chen, Shaomin Zhang, Maomao Zhang, *Member*, IEEE.

*Abstract*—Tumor Treating Fields (TTFields) is a non-invasive anticancer modality that utilizes alternating electric fields to disrupt cancer cell division and growth. While generally well-tolerated with minimal side effects, traditional TTFields therapy for lung tumors faces challenges due to the influence of respiratory motion. We design a novel closed-loop TTFields strategy for lung tumors by incorporating electrical impedance tomography (EIT) for real-time respiratory phase monitoring and dynamic parameter adjustments. Furthermore, we conduct theoretical analysis to evaluate the performance of the proposed method using the lung motion model. Compared to conventional TTFields settings, we observed that variations in the electrical conductivity of lung during different respiratory phases led to a decrease in the average electric field intensity within lung tumors, transitioning from end-expiratory (1.08 V/cm) to end-inspiratory (0.87 V/cm) phases. Utilizing our proposed closed-Loop TTFields approach at the same dose setting (2400 mA, consistent with the traditional TTFields setting), we can achieve a higher and consistent average electric field strength at the tumor site (1.30 V/cm) across different respiratory stages. Our proposed closed-loop TTFields method has the potential to improved lung tumor therapy by mitigating the impact of respiratory motion.

## I. INTRODUCTION

Tumor Treating Fields (TTFields), as a non-invasive approach that uses low-intensity (1-3 V/cm) and intermediate-frequency (100-300 kHz) alternating electric fields, have emerged as a promising therapeutic modality for various malignancies [1]. This treatment strategy harnesses the power of electric fields to disrupt the division of cancer cells, impeding their growth and progression [2, 3]. It is generally well-tolerated by patients and has minimal reported side effects compared to traditional cancer treatments [4].

The spatial distribution of the electric field (EF) generated by TTFields within the tumor areas is a key factor influencing treatment effectiveness [5]. However, directly measuring these spatial EF distributions in live human tissues during TTFields is challenging [6]. Consequently, computational models of TTFields have emerged as an alternative approach to estimate EF distributions and optimize treatment parameters [7, 8]. Each patient has a distinct anatomical structure and unique tumor characteristics, including size, location, and shape. TTFields modeling provides a means for accurately computing the EF distribution in individuals [9-12]. This enables healthcare practitioners to finely adjust TTFields therapy, thereby ensuring the precise targeting of electric fields to the tumor while concurrently minimizing exposure to healthy tissue [9, 13]. The effectiveness of optimizing TTFields therapy, particularly in the treatment of specific cancers like glioblastoma multiforme, has been substantiated in clinical trials [14]. Nevertheless, the optimal application of TTFields encounters challenges when dealing with lung tumors due to the substantial obstacle posed by respiratory motion, which hinders therapeutic efficacy [15].

In the management of TTFields for lung cancer, the respiratory cycle introduces dynamic changes in the shape and electrical properties of lung tissue [16]. This presents a significant challenge to delivering consistent therapeutic electric fields. These fluctuations impact the treatment field within the tumor, potentially reducing treatment effectiveness and leading to suboptimal outcomes. Addressing this limitation has spurred the development of innovative solutions to adapt TTFields treatment to the dynamic nature of lung tumors and their interaction with respiratory motion. The current approaches proposed to address the impact of respiratory motion on tumor treatment include methods such as implanting fiducial markers [17] and respiratory gating [18] which are invasive. Non-invasive methods involve establishing biophysical models to simulate respiratory motion for tracking tumors [19, 20], but these methods often struggle to meet the requirements for real-time therapy. Electrical impedance tomography (EIT) is a non-invasive bedside imaging technique that has been successfully utilized to assess lung function, diagnose respiratory conditions and monitor the patients during mechanical ventilation [21, 22]. EIT can provide real-time monitoring of respiratory motion and has been applied for assessing lung ventilation in patients with respiratory diseases [22, 23]. Additionally, it holds potential for detecting lung cancer and its metastases [24]. In contrast to traditional methods of respiratory phase detection, such as movement, breath sounds, or pulse wave analysis, EIT primarily utilizes surface electrodes to detect signals and assess respiratory status with a relatively high level of accuracy. These electrodes can also be employed alongside TTFields transducers. Therefore, the real-time monitoring of respiratory phases using EIT may be a promising and effective approach to guide TTFields in lung cancer treatment.

The main goal of this study is to introduce a novel TTFields strategy for lung tumors, mitigating the impact of respiratory motion. This innovative approach ensures dependable and efficient treatment field delivery by incorporating real-time respiratory phase monitoring and dynamic adjustment parameters of TTFields, effectively addressing the challenges posed by respiratory motion.

## II. Materials and Methods

### A. Overview of Closed-loop TTFields

As illustrated in Fig. 1, a fundamental component of our proposed closed-loop TTFields method is the incorporation of real-time respiratory phase monitoring and dynamic parameter optimization into the ongoing treatment regimen. This method allows for adaptive adjustment of TTFields parameters based on individual respiratory phase monitoring, thereby enhancing treatment efficacy. For respiratory phase monitoring, we utilized EIT to real-time measure regional lung ventilation distribution by calculating the conductivity changes in the corresponding regions. The measurement electrodes used for EIT can be repurposed for TTFields, thereby ensuring practical convenience. As patients undergo TTFields, real-time EIT data continually updates the current respiratory phase. Subsequently, the dynamic optimization algorithm of TTFields utilizes this information to adjust and refine parameters in response to morphological changes in the lung. Essential data for electrode optimization are sourced from individualized lung motion and induced electric field modeling derived from pre-CT scans. The construction process of the lung motion model and TTFields model are completed offline prior to treatment.

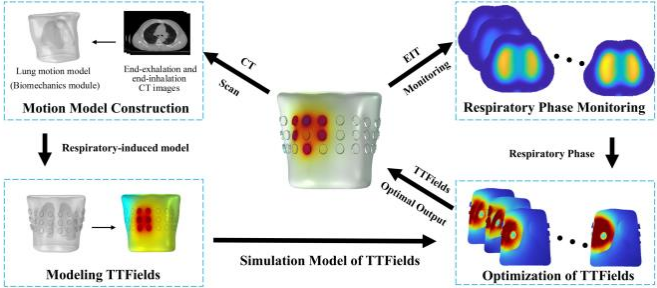

**Fig. 1.** Illustration of adaptive closed-loop TTFields.

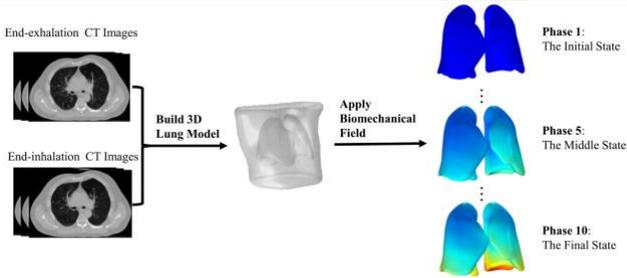

**Fig. 2.** Construction of lung motion model.

### B. Simulation Model in Lung Tumor

#### 1) Construction of the Respiratory-induced Lung Motion Model

The detailed construction process of lung motion model has been elaborated in our previous research [25]. To evaluate the effectiveness of our proposed method, we utilized CT data sourced from 4D Lung Imaging of NSCLC Patients (4D-Lung) [26]. These CT data encompass respiration-correlated scans and were meticulously segmented using 3D Slicer software [27]. For simplicity, the motion model is specifically limited to the chest and lungs, and the lung is represented as a linear, homogeneous, and isotropic elastic tissue. As shown in Fig. 2, we developed a lung motion model that simulates the entire dynamic lung motion during inhalation, utilizing the CT images captured at both end-exhalation to end-inhalation moments. This comprehensive model was employed in subsequent study.

Utilizing the lung motion model, we can derive the structural variations in the lung across different respiratory phases. To facilitate computation, we selected 10 respiratory phases ranging from the end of inhalation to the end of exhalation for EF modeling analysis.

#### 2) Construction of the Induced EF Modeling of TTFields

As depicted in Fig. 3A-B, we created a virtual lung tumor within the lung. This virtual tumor comprises a tumor core and a surrounding tumor shell. The tumor lesions were characterized by an outer radius of 10 mm and an inner core radius of 7 mm, defining a central necrotic core. A total of forty-eight candidate TTFields transducers, each with a 20 mm diameter, are meticulously arranged in a circular fashion on the surface of the chest cavity. These electrodes are organized into upper, middle, and lower sections.

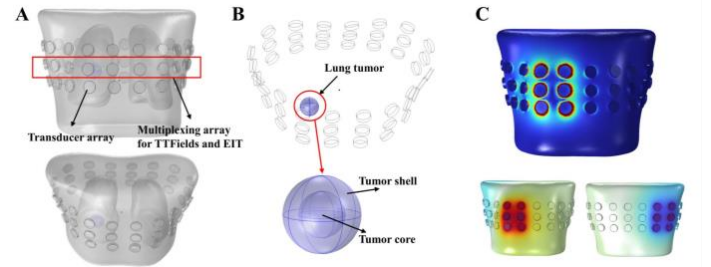

**Fig. 3.** Simulation model of TTFields. A. The image represents the position setting of TTF candidate transducers, with a total of three rounds of transducers in the top, middle, and bottom. Each round has a total of 16 transducer candidate positions, totaling 48 transducer candidate positions. Among them, the 16 transducers in the middle serve as both TTFields and respiratory phase detection using EIT; B. The image represents the relative position of the tumor relative to the TTFields transducer; C. The image represents a case study of electric field simulation results.

TABLE I
DIELECTRIC TISSUE PROPERTIES

| Tissue | Conductivity [S/m] | Relative Permittivity |
|---|---|---|
| Chest | 0.5476 | 825.98 |
| End-exhalation lung tissue | 0.1418 | 3279.6 |
| End-inhalation lung tissue | 0.0622 | 1664.1 |
| Tumor core | 1 | 110 |
| Tumor shell | 0.24 | 2000 |
| Gel | 4.5 | 100 |
| Transducer arrays | 0.0001 | 16000 |

The induced electromagnetic wave lengths of TTFields in the modeled biological tissues are much larger than the size of the Chest. Thus Quasi-static Maxwell's equation is used in our modeling [28]. Consider human body as a homogeneous volume conductor $\Omega$, In the steady state, the electric potential distribution inside model $\Omega$ is governed by the complex quasi-static Laplace equation:

$$\nabla \cdot (\tilde{\sigma} \nabla V) = 0 \quad on\ \Omega \quad (1)$$

$$\tilde{\sigma} = \sigma + i\omega\varepsilon \quad (2)$$

Here, $\tilde{\sigma}$ and $V$ is the complex conductivity and the electrical potential in $\Omega$, respectively. $\sigma$ is the electrical conductivity, $\varepsilon$ is the permittivity.

The induced electric field E was derived from the scalar potential as $E = -\nabla V$, and the current density $J$ was calculated from the electric field using Ohm's law as $J = \tilde{\sigma} E$. The dielectric characteristics of various components are detailed in TABLE I [29, 30].

We calculated the spatial distribution of electric potential using finite element method (FEM) at frequency domain. The FEM solver was implemented using COMSOL Multiphysics.

### C. Electrical Impedance Tomography Measurements for Respiratory Monitoring

In our study, lung EIT systems utilize a 16-electrode sensor evenly distributed around the circumference of the chest, which is placed between the 4th and 6th intercostal space as shown in Fig. 3A [31]. These electrodes are connected to a data acquisition system, which applies a small electrical current through pairs of electrodes and measures the resulting voltage distribution. By analyzing the changes in impedance, the system can monitor the regional ventilation and the changes in lung volume. One complete scan consists of 104 voltage measurements in the adjacent protocol excitation strategy is adopted in this study[32], and the measurements at phase 1 and 10 are plotted in Fig. 4.

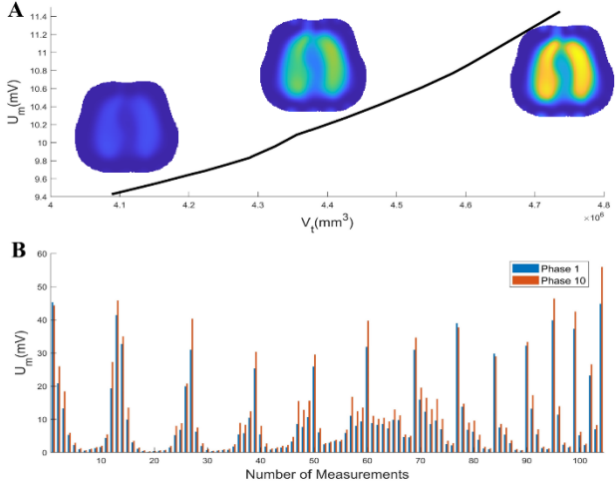

**Fig. 4.** A. The mean value of voltage measurement changes with the lung volume increases, and three and the EIT image reconstructions at phase 1, 8 and phase 10; B. The voltage measurement of 104 channels for a 16-electrode EIT system at respiratory phase 1 and 10.

#### 1) EIT Forward Model

To solve the forward model of EIT, the complete electrode model is used in our study, which can be expressed as:

$$\nabla \cdot \big(\tilde{\sigma}(p)\nabla \phi(p)\big) = 0, p \in \Omega \quad (3)$$

$$\phi(p) + z_l \tilde{\sigma}(p) \frac{\partial \phi(p)}{\partial n} = U_l, p \in e_l, l = 1,2, \dots, L \quad (4)$$

$$\tilde{\sigma}(p) \frac{\partial \phi(p)}{\partial n} = 0, p \in \partial\Omega \setminus \bigcup_{l=1}^{L} e_l \quad (5)$$

$$\int_{e_l} \tilde{\sigma}(p) \frac{\partial \phi(p)}{\partial n} = I_l \quad (6)$$

$$\sum_{l=1}^{L} I_l = 0, \sum_{l=1}^{L} U_l = 0 \quad (7)$$

where the $p$ is the point inside the sensing area $\Omega$; $\phi(p)$ and $\tilde{\sigma}(p)$ denote the potential and complex conductivity at $p$ respectively; $z_l$ is the contact impedance between the electrodes and the body; $n$ is outward unit normal vector to $\partial\Omega$; $U_l$ and $I_l$ is the electrical potential and injected current on the electrode $e_l$; $L$ is the number of electrodes.

#### 2) EIT for Respiratory Monitoring

For image reconstruction of EIT, the distribution of conductivity variation $\dot{G}$ can be reconstructed by solving the inverse problem of EIT. This is usually formulated as an optimization problem:

$$\min_{\dot{G}} \left\{ \big\|\dot{U} - S\dot{G}\big\|^2 + \zeta \cdot l(\dot{G}) \right\} \quad (8)$$

Where $\dot{U}$ is the voltage measurement, S is the sensitivity matrix; $l(\cdot)$ and $\zeta \in R$ denote the regularization function and parameter, respectively. Newton–Raphson (NR) method is used in this study. When $l(\cdot)$ adopts Tikhonov regularization, the NR method can be described as:

$$\hat{\dot{G}}_{k+1} = \hat{\dot{G}}_k - (S^T S + \zeta I)^{-1} S^T \left(S\hat{\dot{G}}_k - \dot{U}\right) \quad (9)$$

We fixed the number of iterations for the NR method at 10 and set the regularization coefficient to $10^4$ in this study. The time taken to reconstruct each image on a PC equipped with MATLAB 2021b, 16GB RAM memory, and an AMD Ryzen 7 6800H CPU is 0.037 seconds.

#### 3) Electrical Properties Setting

In our study, we partitioned the inhalation process into ten distinct phases. The initial phase, denoted as T1, characterizes the state of end-exhalation, while the final phase, denoted as T10, represents the state of end-inhalation. The excitation current frequency employed in EIT is set at 100 kHz. The electrical properties of the deformed lung varied linearly with the volume change:

$$H_T = H_i + (H_l - H_i) * \frac{V_T - V_i}{V_l - V_i} \quad (10)$$

where $H_T$ is the electrical property (conductivity and permittivity) at phase T. $H_i$ and $H_l$ are the electrical properties at the end-exhalation phase and the end-inhalation phase respectively. And $V_T$ is the volume of the deformed lung at phase T.

In Fig. 4B, the mean value of 104 voltage measurement (Um) from EIT changes during the respiration clearly. With continuous inhalation, the electrical impedance of the lung regions increases, leading to an increment in the measured voltage of EIT. This phenomenon exhibits a notable monotonicity. By reconstructing EIT images from the multiple-channel voltage signals, we can obtain informative images that reflect the respiratory phase status. These images and measurements can then be utilized to guide the process of TTFields.

### D. Dynamic Optimization of TTFields by Individualized Lung Simulation Models

The optimization process involves defining a set of parameters for each respiratory phase within the lung simulation models. These parameters are iteratively adjusted to attain optimal TTFields distribution for each phase. The iterative nature of this optimization process ensures that TTFields parameters are dynamically adapted, facilitating consistent field coverage across varying tumor positions. The goal is to ensure that TTFields exert maximal therapeutic impact while minimizing skin reaction and optimizing energy deliver.

*1) Acquisition of the Leadfield Matrix A*

Before optimization, it is necessary to obtain the leadfield matrix A that maps input current to induced EF intensity. leadfield matrix A should be obtained respectively for each phase of lung motion model.

$$e = As \quad (11)$$

where

$$e = [e(r_1)\ e(r_2)\ \cdots\ e(r_N)]^T, \quad (12)$$

$$A = \begin{bmatrix} a_1(r_1) & a_2(r_1) & \cdots & a_M(r_1) \\ a_1(r_2) & a_2(r_2) & \cdots & a_M(r_2) \\ \vdots & \vdots & \cdots & \vdots \\ a_1(r_N) & a_2(r_N) & \cdots & a_M(r_N) \end{bmatrix}, \quad (13)$$

$$s = [s_1\ s_2\ \cdots\ s_M]^T. \quad (14)$$

In the equation above, $e$ represents the vector of EF intensity, $s$ represents the magnitude of the current of the electrodes, $N$ is the number of nodes in the lung model, $M$ is the total number of electrodes, and $r_N$ represents the position of the $n^{th}$ node.

*2) Optimization Method of TTFields*

The l of optimization is to enhance the induced EF within lung tumors for each phase while maintaining a certain degree of EF focality. The overall optimization function is defined as follows:

$$\arg\max_s \left( sum(Cs) - \lambda \cdot sum(e_{0.25max}) \right) \quad (15)$$

where $C$ represents the component leadfield matrix $A$ corresponding to the node where the tumor is located, and $C_s$ represents the EF intensity at the nodes within the tumor. $e_{0.25max}$ denotes the EF intensity at the nodes within a distance of one-fourth of the maximum distance in the model from the center of tumor. $\lambda$ is a weighting parameter used to balance the objectives of EF intensity and focality. A smaller value of $\lambda$ results in higher EF intensity at the location of tumor, while a larger value of $\lambda$ leads to better focality of the EF distribution throughout the lung and chest. Adjusting the value of $\lambda$ allows us to attain the desired optimization outcomes for both EF intensity and focality.

*3) Dose Constraints*

For safety and practical considerations, it is necessary to add some constraints in the optimization process. Firstly, the total current input should be equal to the total current output, meaning the sum of currents should be zero.

$$\sum s_i = 0 \quad (16)$$

Secondly, the total absolute values of the current of all electrodes should not exceed $I_{total}$.

$$\sum |s_i| \leq I_{total} \quad (17)$$

Thirdly, the absolute values of the current of each individual electrode should not exceed $I_{max}$.

$$|s_i| \leq I_{\max} \quad (18)$$

In addition, constraints can be imposed on the number of electrodes used to facilitate practical application, such as using a fixed number of n electrodes.

$$Electrode\ number = n \quad (19)$$

### E. Evaluation of Closed-loop TTFields

To assess the efficacy of our novel approach, we established a control setting employing traditional TTFields parameters, as depicted in Figure 3C. Traditional TTFields consisted of 12 intervention electrodes, strategically positioned in alignment with the location of lung tumor. Each electrode administered an input current of 200 mA, resulting in a total dose of 2400 mA (Traditional setting). Concurrently, we configured two closed-loop TTFields settings. One setting (Same dose setting) underwent optimization based on the total dose (2400 mA) administered in the control group. The other one (Different dose setting) was optimized with the target of achieving an average EF intensity of 1.0 V/cm at the tumor shell, subsequently determining the requisite total dose. We conducted EF simulations for these three distinct TTFields settings, each corresponding to different respiratory phases. Subsequently, we compared the electric field intensity within the tumor at various respiratory phases.

## III. RESULTS

As shown in Fig. 5, we observed that within the traditional setting of TTFields, variations in the electrical conductivity of lung during different respiratory phases led to a decrease in the average electric field intensity within lung tumors, transitioning from end-expiratory (1.08 V/cm) to end-inspiratory (0.87 V/cm) phases. However, employing our proposed dynamic optimization approach with the Same dose setting (The total

## IV. Discussion

Our proposed closed-loop TTFields method represents an innovative approach to treatment lung tumors. Different from conventional strategy [33], our closed-loop TTFields dynamically adjust in response to real-time respiratory phases monitoring. Quantitative assessments reveal that the adaptive closed-loop method consistently maintains optimal TTFields coverage throughout different respiratory phases. Dynamic adjustments of TTFields parameters in response to changing tumor positions effectively mitigate the influence of respiratory motion. Comparison with traditional TTFields delivery methods highlights the superior performance of the closed-loop feedback approach in preserving uniform electric field distribution across the tumor volume.

A pivotal aspect of our closed-loop TTFields method revolves around the intricate field of dynamic lung modeling. Constructing personalized lung simulation models requires careful consideration of the intricate and multifaceted aspects of lung physiology. These models encapsulate the dynamic interplay between lung tissue and tumor motion during respiration[16]. The rhythmic expansion and contraction of the lung during breathing lead to variations in tumor position and shape. These fluctuations introduce complexity into treatment delivery, challenging the uniformity of therapeutic electric fields. Through dynamic lung modeling, we confront this complexity head-on, capturing the nuanced variations in lung expansion and tumor movement. Consequently, our closed-loop system is empowered to adapt and fine-tune TTFields in real time, resulting in a notable enhancement in therapy precision.

The computational complexity of dynamic modeling cannot overlook. The task of continuously updating and refining models in real time demands computational resources and algorithms that can keep pace. Thus, it becomes imperative to strike a balance between model complexity and computational feasibility. Our algorithm engages in preparatory procedures, predominantly via pre-computation, thereby endowing it with the capability to expeditiously effectuate real-time adaptations during therapeutic sessions. This rapid responsiveness assumes pivotal significance, for it guarantees the meticulous synchronization of our treatment protocol with the dynamically fluctuating conditions within the lung throughout the entire therapeutic session. Furthermore, our algorithm adeptly manages the equilibrium between focality and intensity of induced EF, taking into full account the geometric structure of the individualized lung. Through the optimization of individual parameters, we can enhance the focality and intensity of the induced EF, thereby contributing to enhanced treatment outcomes.

In this study, a specific total intervention dose was employed for the optimization of TTFields. It is noteworthy that through individual optimization strategies, we observed that compared to traditional TTFields parameter settings, we could achieve a higher therapeutic field strength in the tumor with the same total current (2400 mA). Furthermore, when setting the preset field strength at the tumor site to 1V/cm (a commonly used threshold dose was the same as Traditional setting, 2400 mA) allowed us to achieve a more extensive distribution of the treatment field (1.30 V/cm), maintaining consistency within the tumor across diverse respiratory phases. The optimized transducer settings are detailed in Supplementary material.

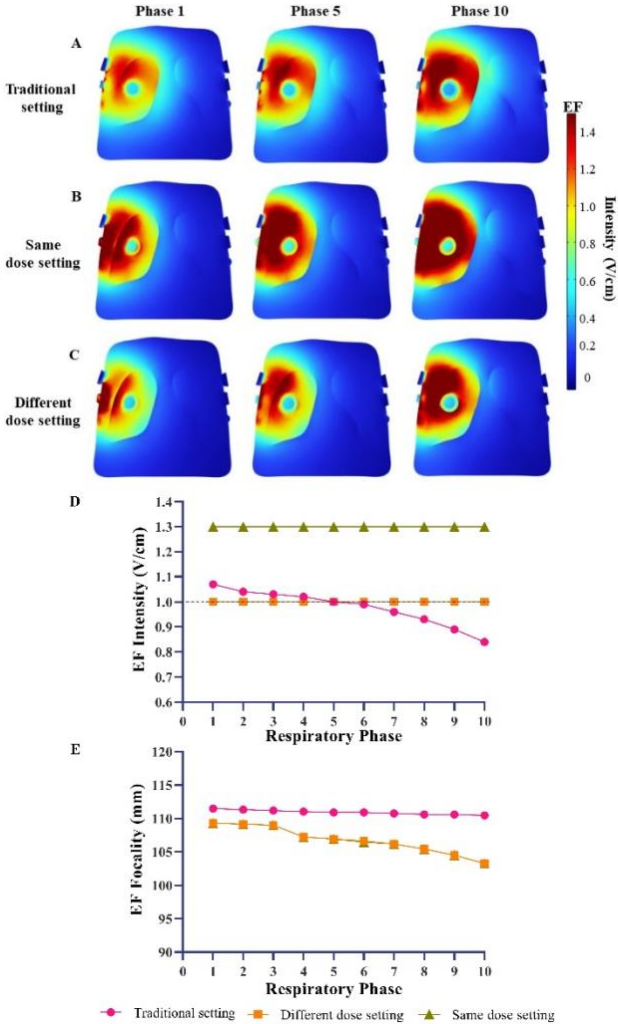

**Fig. 5.** The induced EF distribution from different TTFields setting at different respiratory phases. A-C. The induced EF distribution at different respiratory phases for different TTFields setting (Traditional setting, Same dose setting, Different dose setting). D. Distribution of mean electric field intensity generated within tumor shell by different TTFields settings. E. Focality of electric fields generated within tumor shell under different TTFields settings. Phase 1(end-expiratory), Phase 10 (end-inspiratory).

Furthermore, when we maintained an average treatment field intensity of 1 V/cm at the tumor shell with dynamic TTFields optimization (Different dose setting), we were able to achieve a 23.10% reduction (1338 mA) in the required total output dose compared to the Traditional setting. This reduction in dose was achieved while ensuring uniform treatment field distribution within the tumor, demonstrating the efficacy of our approach. Simultaneously, our dynamic optimization method enhances the focality of electric fields generated by TTFields within brain compared to the traditional TTFields setting. The majority of the intervention current energy is concentrated around the tumor site, particularly during the gradual inhalation process.

for therapeutic field efficacy in TTFields research), we found that the total current demand could be reduced by 23.10%, indicating that individual optimization can lower the total dose requirement. The prevalence of skin reactions beneath the transducer arrays constituted the most frequently documented adverse events in clinical reports [34]. The dose optimization of TTFields is imperative for safety and practical considerations, thus the integration of temperature field simulations and measurements becomes essential [35]. This integration serves the crucial purpose of mitigating potential temperature-related side effects induced by TTFields [36]. By carefully managing the interplay between TTFields and temperature fields, practitioners can ensure that therapeutic outcomes are maximized while minimizing any undesirable thermal effects, thus enhancing the safety and efficacy of this innovative approach.

EIT, as a non-invasive and radiation-free measurement technique, has been widely applied in clinical settings to achieve functional imaging of lung ventilation in patients. In traditional tumor radiotherapy, patients are typically instructed to hold their breath to minimize tumor motion, allowing radiation doses to be delivered specifically to the tumor region within a relatively short treatment process. However, TTFields usually require patients to wear stimulating electrodes for about 20 hours each day to effectively disrupt cancer cell division and growth, and patients cannot maintain a static breath-hold during this wearing period. Therefore, the utilization of EIT technology for monitoring respiratory phases significantly enhances the effectiveness of TTFields treatment. Another significant advantage of this closed-loop method is the reuse of electrodes between TTFields and EIT. The electrodes distributed around the thoracic cavity for generating the therapeutic electric field in TTFields application can conveniently serve as measurement electrodes for EIT, enabling the implementation of closed-loop TTFields without the need for additional sensors.

The closed-loop TTFields method exemplifies the convergence of technology and personalized medicine, holding the potential to reshape the landscape of lung tumor therapy. As technology continues to evolve, the realization of a dynamic, patient-centric approach to cancer treatment draws closer, offering renewed hope for improved therapeutic outcomes and enhanced life quality for lung cancer patients. Looking ahead, further research is imperative to refine the clinical applicability of this method. Extended clinical investigations will yield thorough insights into how the approach influences treatment responses and patient outcomes. Investigations into refining individualized lung simulation models and optimizing the closed-loop feedback algorithm are essential. Additionally, the integration of EIT data into the closed-loop feedback algorithm demands a meticulous approach to data preprocessing and noise reduction to uphold the fidelity of the adaptive treatment process.

A pragmatic perspective necessitates acknowledging the limitations of this study. In our study, two simplified models are used: the first one is lung geometric model without detailed tissue segmentation to create a more refined model, and the second one is lung motion model without the movement of outer contour of the chest to consider the movement of both the electrodes and chest space outside the lungs. These simplifications could affect the accuracy and realism of the model. Furthermore, the dynamic fluctuations in tumor position and shape and microscale modeling of lung alveoli during respiratory motion did not account. There is potential to investigate more intricate lung and tumor models that can provide a more accurate simulation of tissue complexities in future research [16]. Additionally, considering the anatomical proximity of the lungs to the heart, the impacts of cardiac-induced tissue deformation and its subsequent influence on tumor vascular supply modifications could be more precisely delineated through the amalgamation of electrocardiographic monitoring with the proposed methodological framework. It should be noted that our method has not been applied or validated in actual patients. While our research proposes a promising approach, its practical effectiveness has not been verified. Further clinical studies and real-world applications are necessary to determine the feasibility and efficacy of this method in real patients.

## V. Conclusion

By implementing our proposed closed-loop TTFields approach at the same dose setting (2400 mA, aligning with traditional TTFields settings), we have achieved a notably higher and consistent average electric field strength at the tumor site (1.30 V/cm) across various respiratory stages. Additionally, through dynamic TTFields optimization to maintain an average treatment field intensity of 1 V/cm at the tumor shell (using the different dose setting), we observed a substantial 23.10% reduction (1338 mA) in the required total output dose compared to the traditional setting. These findings suggest that our closed-loop TTFields method holds promise for advancing lung tumor therapy by effectively addressing the challenges posed by respiratory motion.

## References

[1] C. Wenger et al., "A review on tumor-treating fields (TTFields): clinical implications inferred from computational modeling," IEEE Reviews in Biomedical Engineering, vol. 11, pp. 195-207, 2018.
[2] E. D. Kirson et al., "Alternating electric fields arrest cell proliferation in animal tumor models and human brain tumors," Proceedings of the National Academy of Sciences, vol. 104, no. 24, p. 10152, 2007.
[3] N. K. Karanam and M. D. Story, "An overview of potential novel mechanisms of action underlying Tumor Treating Fields-induced cancer cell death and their clinical implications," Int. J. Radiat. Biol., vol. 97, no. 8, pp. 1044-1054, 2021/08/03 2021.
[4] M. Pless and U. Weinberg, "Tumor treating fields: concept, evidence and future," Expert Opinion on Investigational Drugs, vol. 20, no. 8, pp. 1099-1106, 2011/08/01 2011.
[5] E. D. Kirson et al., "Disruption of Cancer Cell Replication by Alternating Electric Fields," Cancer Res., vol. 64, no. 9, pp. 3288-3295, 2004.
[6] M. Ma et al., "Validation of Computational Simulation for Tumor-treating Fields with Homogeneous Phantom," in 2022 44th Annual International Conference of the IEEE Engineering in Medicine & Biology Society (EMBC), 2022, pp. 975-978.

[7] P. C. Miranda, A. Mekonnen, R. Salvador, and P. J. Basser, "Predicting the electric field distribution in the brain for the treatment of glioblastoma," Phys. Med. Biol., vol. 59, no. 15, pp. 4137-4147, 2014/07/08 2014.

[8] C. Wenger, R. Salvador, P. J. Basser, and P. C. Miranda, "Improving Tumor Treating Fields Treatment Efficacy in Patients With Glioblastoma Using Personalized Array Layouts," International Journal of Radiation Oncology*Biology*Physics, vol. 94, no. 5, pp. 1137-1143, 2016.

[9] A. R. Korshoej, J. C. H. Sørensen, G. von Oettingen, F. R. Poulsen, and A. Thielscher, "Optimization of tumor treating fields using singular value decomposition and minimization of field anisotropy," Phys. Med. Biol., vol. 64, no. 4, p. 04NT03, 2019/02/08 2019.

[10] C. Wenger, R. Salvador, P. J. Basser, and P. C. Miranda, "The electric field distribution in the brain during TTFields therapy and its dependence on tissue dielectric properties and anatomy: a computational study," Phys. Med. Biol., vol. 60, no. 18, pp. 7339-7357, 2015/09/09 2015.

[11] E. Lok, O. Liang, T. Malik, and E. T. Wong, "Computational Analysis of Tumor Treating Fields for Non-Small Cell Lung Cancer in Full Thoracic Models," Advances in Radiation Oncology, vol. 8, no. 4, p. 101203, 2023/07/01/ 2023.

[12] A. R. Korshoej, F. L. Hansen, N. Mikic, G. von Oettingen, J. C. H. Sørensen, and A. Thielscher, "Importance of electrode position for the distribution of tumor treating fields (TTFields) in a human brain. Identification of effective layouts through systematic analysis of array positions for multiple tumor locations," PLoS One, vol. 13, no. 8, p. e0201957, 2018.

[13] N. Gentilal, A. Naveh, T. Marciano, and P. C. Miranda, "The Impact of Scalp's Temperature on the Choice of the Best Layout for TTFields Treatment," IRBM, vol. 44, no. 3, p. 100768, 2023/06/01/ 2023.

[14] C. Straube et al., "Dosimetric impact of tumor treating field (TTField) transducer arrays onto treatment plans for glioblastomas–a planning study," Radiation Oncology, vol. 13, no. 1, p. 31, 2018.

[15] Z. Bomzon et al., "Using computational phantoms to improve delivery of Tumor Treating Fields (TTFields) to patients," in 2016 38th Annual International Conference of the IEEE Engineering in Medicine and Biology Society (EMBC), 2016, pp. 6461-6464.

[16] T. Zhou, S. Liu, H. Lu, J. Bai, L. Zhi, and Q. Shi, "Nested multi-scale transform fusion model: The response evaluation of chemoradiotherapy for patients with lung tumors," Comput. Methods Programs Biomed., vol. 232, p. 107445, 2023/04/01/ 2023.

[17] H. Shirato et al., "Speed and amplitude of lung tumor motion precisely detected in four-dimensional setup and in real-time tumor-tracking radiotherapy," International Journal of Radiation Oncology*Biology*Physics, vol. 64, no. 4, pp. 1229-1236, 2006/03/15/ 2006.

[18] M. Mizuhata et al., "Respiratory-Gated Proton Beam Therapy for Hepatocellular Carcinoma Adjacent to the Gastrointestinal Tract without Fiducial Markers," Cancers (Basel), vol. 10, no. 2, p. 58, 2018.

[19] R. Werner, J. Ehrhardt, R. Schmidt, and H. Handels, Modeling respiratory lung motion: a biophysical approach using finite element methods (Medical Imaging). SPIE, 2008.

[20] H. Ladjal, M. Beuve, P. Giraud, and B. Shariat, "Towards Non-Invasive Lung Tumor Tracking Based on Patient Specific Model of Respiratory System," IEEE Trans. Biomed. Eng., vol. 68, no. 9, pp. 2730-2740, 2021.

[21] N. Kashibe, F. Fujii, T. Shiinoki, and K. Shibuya, "Construction of a respiratory-induced lung tumor motion model using phase oscillator," in 2017 IEEE International Conference on Systems, Man, and Cybernetics (SMC), 2017, pp. 699-704.

[22] V. Tomicic and R. Cornejo, "Lung monitoring with electrical impedance tomography: technical considerations and clinical applications," (in eng), J. Thorac. Dis., vol. 11, no. 7, pp. 3122-3135, Jul 2019.

[23] Y. Shi, Z. Yang, F. Xie, S. Ren, and S. Xu, "The Research Progress of Electrical Impedance Tomography for Lung Monitoring," (in English), Frontiers in Bioengineering and Biotechnology, Review vol. 9, 2021-October-01 2021.

[24] B. Sun, S. Yue, Z. Hao, Z. Cui, and H. Wang, "An Improved Tikhonov Regularization Method for Lung Cancer Monitoring Using Electrical Impedance Tomography," IEEE Sensors Journal, vol. 19, no. 8, pp. 3049-3057, 2019.

[25] L. Zhu, W. Lu, M. Soleimani, Z. Li, and M. Zhang, "Electrical Impedance Tomography Guided by Digital Twins and Deep Learning for Lung Monitoring," IEEE Trans. Instrum. Meas., vol. 72, pp. 1-9, 2023.

[26] S. Balik et al., "Evaluation of 4-dimensional Computed Tomography to 4-dimensional Cone-Beam Computed Tomography Deformable Image Registration for Lung Cancer Adaptive Radiation Therapy," International Journal of Radiation Oncology*Biology*Physics, vol. 86, no. 2, pp. 372-379, 2013/06/01/ 2013.

[27] R. Kikinis, S. D. Pieper, and K. G. Vosburgh, "3D Slicer: A Platform for Subject-Specific Image Analysis, Visualization, and Clinical Support," in Intraoperative Imaging and Image-Guided Therapy, F. A. Jolesz, Ed. New York, NY: Springer New York, 2014, pp. 277-289.

[28] E. Lok and E. Sajo, "Fundamental Physics of Tumor Treating Fields," in Alternating Electric Fields Therapy in Oncology: A Practical Guide to Clinical Applications of Tumor Treating Fields, E. T. Wong, Ed. Cham: Springer International Publishing, 2016, pp. 15-27.

[29] E. J. Woo, P. Hua, J. G. Webster, and W. J. Tompkins, "Measuring lung resistivity using electrical impedance tomography," IEEE Trans. Biomed. Eng., vol. 39, no. 7, pp. 756-760, 1992.

[30] Y. Lu, B. Li, J. Xu, and J. Yu, "Dielectric properties of human glioma and surrounding tissue," Int. J. Hyperthermia, vol. 8, no. 6, pp. 755-760, 1992/01/01 1992.

[31] L. Yang et al., "A Wireless, Low-Power, and Miniaturized EIT System for Remote and Long-Term Monitoring of Lung Ventilation in the Isolation Ward of ICU," IEEE Trans. Instrum. Meas., vol. 70, pp. 1-11, 2021.

[32] J. K. Seo, K. C. Kim, A. Jargal, K. Lee, and B. Harrach, "A Learning-Based Method for Solving Ill-Posed Nonlinear Inverse Problems: A Simulation Study of Lung EIT," SIAM Journal on Imaging Sciences, vol. 12, no. 3, pp. 1275-1295, 2019.

[33] Z. Bomzon et al., "Modelling Tumor Treating Fields for the treatment of lung-based tumors," in 2015 37th Annual International Conference of the IEEE Engineering in Medicine and Biology Society (EMBC), 2015, pp. 6888-6891.

[34] S. Liu et al., "Progress and prospect in tumor treating fields treatment of glioblastoma," Biomed. Pharmacother., vol. 141, p. 111810, 2021/09/01/ 2021.

[35] N. Gentilal and P. C. Miranda, "Heat transfer during TTFields treatment: Influence of the uncertainty of the electric and thermal parameters on the predicted temperature distribution," Comput. Methods Programs Biomed., vol. 196, p. 105706, 2020/11/01/ 2020.

[36] N. Gentilal et al., "Temperature and impedance variations during tumor treating fields (TTFields) treatment," Front. Hum. Neurosci., vol. 16, p. 436, 2022.